\documentclass[italian,english]{article}
\usepackage[T1]{fontenc}
\usepackage[latin1]{inputenc}
\usepackage{color}
\usepackage{amssymb}

\makeatletter

 \newcommand{\lyxaddress}[1]{
   \par {\raggedright #1 
   \vspace{1.4em}
   \noindent\par}
 }

\usepackage{babel}
\makeatother
\begin{document}

\title{\textbf{A non-geodesic motion in the $R^{-1}$ theory of gravity
tuned with observations }}

\author{\textbf{Christian Corda}}

\maketitle

\lyxaddress{\begin{center}INFN - Sezione di Pisa and Università di Pisa, Largo
Pontecorvo 3, I - 56127 PISA, Italy; Instituto de Cosmologia, Relatividade
e Astrofìsica (ICRA-BR), Centro Brasilero de Pesquisas Fisicas, Rua
Dr. Xavier Sigaud 150, CEP 22290 -180 Urca Rio de Janeiro - RJ Brazil \end{center}}

\lyxaddress{\begin{center}\textit{E-mail address:} \textcolor{blue}{christian.corda@ego-gw.it}\end{center}}

\begin{abstract}
In the general picture of high order theories of gravity, recently,
the \textbf{$R^{-1}$} theory has been analyzed in two different frameworks.
In this letter a third context is added, considering an explicit coupling
between the \textbf{$R^{-1}$} function of the Ricci scalar and the
matter Lagrangian. The result is a non-geodesic motion of test particles
which, in principle, could be connected with Dark Matter and Pioneer
anomaly problems. 
\end{abstract}

\lyxaddress{PACS numbers: 04.50.+h, 04.20.Jb.}

The accelerated expansion of the Universe, which is today observed,
shows that cosmological dynamic is dominated by the so called Dark
Energy which gives a large negative pressure. This is the standard
picture, in which such new ingredient is considered as a source of
the \textit{right hand side} of the field equations. It should be
some form of un-clustered non-zero vacuum energy which, together with
the clustered Dark Matter, drives the global dynamics. This is the
so called {}``concordance model'' (ACDM) which gives, in agreement
with the Cosmic Microwave Background Radiation (CMBR), Dim Lyman Limit
Systems (LLS) and type la supernovae (SNeIa) data, a good trapestry
of the today observed Universe, but presents several shortcomings
as the well known {}``coincidence'' and {}``cosmological constant''
problems \cite{key-1}. An alternative approach is changing the \textit{left
hand side} of the field equations, seeing if observed cosmic dynamics
can be achieved extending general relativity \cite{key-2,key-3,key-4,key-5,key-6}.
In this different context, it is not required to research candidates
for Dark Energy and Dark Matter, that, till now, have not been found,
but only the {}``observed'' ingredients, which are curvature and
baryonic matter, have to be taken into account. Considering this point
of view, one can think that gravity is not scale-invariant \cite{key-7}
and a room for alternative theories is present \cite{key-8,key-9,key-10,key-26,key-27,key-29}.
In principle, the most popular Dark Energy and Dark Matter models
can be achieved considering $f(R)$ theories of gravity \cite{key-7,key-11},
where $R$ is the Ricci curvature scalar. 

In this picture even the sensitive detectors for gravitational waves,
like bars and interferometers (i.e. those which are currently in operation
and the ones which are in a phase of planning and proposal stages)
\cite{key-12,key-13,key-31,key-33}, could, in principle, be important
to confirm or rule out the physical consistency of general relativity
or of any other theory of gravitation. This is because, in the context
of Extended Theories of Gravity, some differences between General
Relativity and the others theories can be pointed out starting by
the linearized theory of gravity \cite{key-14,key-15,key-16,key-17,key-28,key-32}.

In the general picture of high order theories of gravity, recently,
the \textbf{$R^{-1}$} theory has been analyzed in two different frameworks
\cite{key-15,key-18}. In this letter a third context is added, considering
an explicit coupling between the \textbf{$R^{-1}$} function of the
Ricci scalar and the matter Lagrangian. The result is a non-geodesic
motion of test particles which, in principle, could be connected with
Dark Matter and Pioneer anomaly problems. 

In \cite{key-15} the high order action of the $R^{-1}$ theory of
gravity\begin{equation}
S=\int d^{4}x\sqrt{-g}R^{-1}+\mathcal{L}_{m},\label{eq: high order 1}\end{equation}

has been analyzed in a context of production and potential detection
of gravitational waves, while in \cite{key-18} a cosmological application
of such action has been performed.

Equation (\ref{eq: high order 1}) is a particular choice with respect
to the well known canonical one of general relativity (the Einstein
- Hilbert action \cite{key-19,key-20}) which is 

\begin{equation}
S=\int d^{4}x\sqrt{-g}R+\mathcal{L}_{m}.\label{eq: EH}\end{equation}

Now let us consider a third action including a coupling between the
$R^{-1}$ function of the Ricci scalar and the matter Lagrangian:\begin{equation}
S=\int d^{4}x\sqrt{-g}(R^{-1}+R^{-1}\mathcal{L}_{m}+\mathcal{L}_{m}).\label{eq: high order 2}\end{equation}

Cleary, this feature implies a breakdown of Einstein's equivalence
principle, i.e. a non geodesic motion of test particles \cite{key-21}.

If the gravitational Lagrangian is nonlinear in the curvature invariants,
the Einstein field equations are an order higher than second \cite{key-5,key-6,key-8,key-14,key-15}.
For this reason such theories are often called higher-order gravitational
theories. This is exactly the case of the action (\ref{eq: high order 2}).

If one varies this action with respect to $g_{\mu\nu}$ (see also
refs. \cite{key-8,key-14,key-15} for a parallel computation) the
field equations are obtained (note that in this paper we work with
$G=1$, $c=1$ and $\hbar=1$):

\begin{equation}
\begin{array}{c}
-R^{-2}R_{\mu\nu}-\frac{R^{-1}}{2}g_{\mu\nu}+\bigtriangledown_{\mu}\bigtriangledown_{\nu}R^{-2}-g_{\mu\nu}\square R^{-2}=\\
\\=2R^{-2}\mathcal{L}_{m}R_{\mu\nu}-2(\bigtriangledown_{\mu}\bigtriangledown_{\nu}-g_{\mu\nu}\square)R^{-2}\mathcal{L}_{m}+(1+R^{-1})T_{\mu\nu}^{(m)},\end{array}\label{eq: einstein-general}\end{equation}

where $\square$ is the d' Alembertian operator and $T_{\mu\nu}^{(m)}$
is the ordinary stress-energy tensor of the matter. 

Following the analysis in \cite{key-22}, let us compute the covariant
derivative of the field equations (\ref{eq: einstein-general}) ,
together with the Bianchi identity \cite{key-19} 

\begin{equation}
\bigtriangledown^{\mu}G_{\mu\nu}=0\label{eq: Id Bianchi}\end{equation}

and with the identity 

\begin{equation}
(\square\bigtriangledown_{\nu}-\bigtriangledown_{\nu}\square)R^{-2}=R_{\mu\nu}\bigtriangledown^{\mu}R^{-2},\label{eq: Identity}\end{equation}

obtaining \begin{equation}
\bigtriangledown^{\mu}T_{\mu\nu}^{(m)}=\frac{-R^{-2}}{R^{-1}+1}(g_{\mu\nu}\mathcal{L}_{m}-T_{\mu\nu}^{(m)})\bigtriangledown^{\mu}R.\label{eq: TEI}\end{equation}

From the last equation it seems that energy conservation breaks down
\cite{key-21}. This apparent shortcoming can be avoided in the following
way. Writing down, explicitly, the Einstein tensor in equation (\ref{eq: einstein-general})
one gets

\begin{equation}
\begin{array}{c}
G_{\mu\nu}=-R^{2}\bigtriangledown_{\mu}\bigtriangledown_{\nu}R^{-2}+R^{2}g_{\mu\nu}\square R^{-2}+Rg_{\mu\nu}\\
\\+2\mathcal{L}_{m}R_{\mu\nu}-2R^{2}(\bigtriangledown_{\mu}\bigtriangledown_{\nu}-g_{\mu\nu}\square)R^{-2}\mathcal{L}_{m}+(R^{2}+R)T_{\mu\nu}^{(m)}.\end{array}\label{eq: einstein-general 2}\end{equation}

Then, one can introduce a {}``total'' stress-energy tensor \begin{equation}
\begin{array}{c}
T_{\mu\nu}^{(tot)}\equiv-\frac{1}{8\pi}\{ R^{2}\bigtriangledown_{\mu}\bigtriangledown_{\nu}R^{-2}+R^{2}g_{\mu\nu}\square R^{-2}+Rg_{\mu\nu}\\
\\+2\mathcal{L}_{m}R_{\mu\nu}-2R^{2}(\bigtriangledown_{\mu}\bigtriangledown_{\nu}-g_{\mu\nu}\square)R^{-2}\mathcal{L}_{m}+(R^{2}+R)T_{\mu\nu}^{(m)}\}.\end{array}\label{eq: tensore totale}\end{equation}

In this way the field equations assume the well known Einsteinian
form \begin{equation}
G_{\mu\nu}=8\pi T_{\mu\nu}^{(tot)}.\label{eq: Einstein classiche}\end{equation}

In the {}``total'' stress-energy tensor (\ref{eq: tensore totale})
a \textit{curvature} contribution is added and mixed to the \textit{material}
one. This is because the high order terms contribute, like sources,
to the field equations and can be considered like \textit{effective
fields} (see ref. \cite{key-25} for details).

Thus, using equation (\ref{eq: Einstein classiche}), equation (\ref{eq: Id Bianchi})
gives \begin{equation}
\bigtriangledown^{\mu}T_{\mu\nu}^{(tot)},=0\label{eq: ConsEn}\end{equation}

i.e. the conservation of the total energy (\textit{material} plus
\textit{curvature} plus mixed terms). Moreover, using equation (\ref{eq: tensore totale}),
equation (\ref{eq: ConsEn}) gives directly equation (\ref{eq: TEI}).

With the goal of testing the motion of test particles in the model,
one can introduce the well known stress-energy tensor of a perfect
fluid \cite{key-19,key-20,key-22} \begin{equation}
T_{\mu\nu}^{(m)}\equiv(\epsilon+p)u_{\mu}u_{\nu}-pg_{\mu\nu}.\label{eq: pf}\end{equation}

Because two astrophysical examples will be considered (the first in
the galaxy and the second in the Solar System), this simplest version
of a stress-energy tensor for the matter, which concerns {}``dust''
(i.e. stars at the galactic scale, planets and spacecrafts at the
Solar System scale), can be used in a good approximation \cite{key-19,key-20,key-25}.
In equation (\ref{eq: pf}) $\epsilon$ is the proper energy density,
$p$ the pressure and $u_{\mu}$ the fourth-velocity of the particles.

Now, following \cite{key-22}, one defines the \textit{projector operator}
\begin{equation}
P_{\mu\alpha}\equiv g_{\mu\alpha}-u_{\mu}u_{\alpha},\label{eq: po}\end{equation}

the contraction $g^{\alpha\beta}P_{\mu\beta}$ can be applied to equation
(\ref{eq: TEI}), obtaining \begin{equation}
\frac{d^{2}x^{\alpha}}{ds^{2}}+\tilde{\Gamma}_{\mu\nu}^{\alpha}\frac{dx^{\mu}}{ds}\frac{dx^{\nu}}{ds}=F^{\alpha}.\label{eq: geod}\end{equation}

The presence of the extra force \begin{equation}
F^{\alpha}\equiv(\epsilon+p)^{-1}P^{\alpha\nu}[(\frac{-R^{-2}}{R^{-1}+1})(\mathcal{L}_{m}+p)\bigtriangledown_{\nu}R+\bigtriangledown_{\nu}p]\label{eq: extra}\end{equation}

shows that the motion of test particles is non-geodesic. It is also
simple to see that

\begin{equation}
F^{\alpha}\frac{dx_{\alpha}}{ds}=0,\label{eq: ortogonali}\end{equation}

i.e. the extra force is orthogonal to the four-velocity of test masses.

Taking the Newtonian limit in three dimensions of equation (\ref{eq: geod})
one obtains

\begin{equation}
\overrightarrow{a}_{tot}=\overrightarrow{a}_{n}+\overrightarrow{a}{}_{ng}\label{eq: vect}\end{equation}

where the total acceleration $\overrightarrow{a}_{tot}$ is given
by the ordinary Newtonian acceleration $\overrightarrow{a}_{n}$ plus
the acceleration $\overrightarrow{a}_{ng}$ which is due to the extra
force (non-geodesic).

Using equation (\ref{eq: vect}) and a bit of three-dimensional geometry
the Newtonian acceleration $\overrightarrow{a}_{n}$ can be written
as

\begin{equation}
\overrightarrow{a}_{n}=\frac{1}{2}(a_{tot}^{2}-a_{n}^{2}-a_{ng}^{2})\frac{\overrightarrow{a}_{tot}}{a_{ng}a_{tot}}.\label{eq: Newt}\end{equation}

In the limit in which $\overrightarrow{a}{}_{ng}$ dominates (i.e.
$a_{n}\ll a_{tot}$) it is

\begin{equation}
\label{eq: a tot}\end{equation}
\begin{equation}
a_{n}\simeq\frac{a_{tot}\overrightarrow{a}_{tot}}{2a_{ng}}(1-\frac{a_{ng}^{2}}{a_{tot}^{2}}).\label{eq: a1}\end{equation}

Defining \cite{key-22,key-23,key-24,key-30}

\begin{equation}
a_{e}^{-1}\equiv\frac{1}{2a_{ng}}(1-\frac{a_{ng}^{2}}{a_{tot}^{2}}),\label{eq: aE}\end{equation}

equation (\ref{eq: a1}) becomes \begin{equation}
\overrightarrow{a}_{n}\simeq\frac{a_{tot}}{a_{e}}\overrightarrow{a}_{tot}.\label{eq: a1 due}\end{equation}

From equation (\ref{eq: a1 due}) one gets

\begin{equation}
a_{tot}\simeq(a_{e}\textrm{}a_{n})^{\frac{1}{2}}.\label{eq: a tot}\end{equation}

Because the standard Newtonian acceleration is 

\begin{equation}
\overrightarrow{a}_{n}=\frac{M}{r^{2}}\widehat{u}_{r},\label{eq: StNe}\end{equation}

the total acceleration results\begin{equation}
\overrightarrow{a}_{tot}=\frac{(a_{e}M)^{\frac{1}{2}}}{r}\widehat{u}_{r}=\frac{v_{r}^{2}}{r}\widehat{u}_{r},\label{eq: a tot rad}\end{equation}

where \begin{equation}
v_{r}=(a_{e}M)^{\frac{1}{4}}\label{eq: vr}\end{equation}

is the rotation velocity of a test mass under the influence of the
non-geodesic force.

The extra force has environmental nature, thus only phenomenology
can help us in its identification.

In a galactic context it is natural to identify $a_{e}$ with $a_{0}\simeq10^{-10}m/s$,
which is the acceleration of Milgrom used in the theoretical context
of Modified Newtonian Dynamics to achieve Dark Matter into galaxies
\cite{key-22,key-24}. 

From another point of view, in the Solar System, if the anomaly in
Pioneer acceleration is not generated by systematic effects, but a
real effect is present \cite{key-22,key-24}, one can in principle
put

\begin{equation}
a_{e}=a_{Pi}\simeq8.5\times10^{-10}m/s^{2}.\label{eq: aP}\end{equation}

Thus, the introduced approach allows for a unified explanation of
the two effects.

\section*{Conclusions}

In this letter the \textbf{$R^{-1}$} theory of gravity has been analyzed
in a third context which has been added to two different frameworks
recently seen in the general picture of high order theories of gravity.
An explicit coupling between the \textbf{$R^{-1}$} function of the
Ricci scalar and the matter Lagrangian has been performed. The result
is a non-geodesic motion of test particles which, in principle, could
be connected with Dark Matter and Pioneer anomaly problems.

\section*{Acknowledgements }

I would like to thank Salvatore Capozziello, Mauro Francaviglia and
Maria Felicia De Laurentis for useful discussions on the topics of
this paper. I thank the referee for precious advices and comments
that allowed to improve this letter, for the physical significance
and for the style too. The EGO consortium has also to be thanked for
the use of computing facilities. This research has been partially
supported by the ICRA-Brazil.

\end{document}